\documentclass[10pt,twoside,notitlepage,twocolumn]{article}
\usepackage{letter}

\newcommand\email[1]{%
	\begingroup
	\renewcommand\thefootnote{}\footnote{#1}%
	\addtocounter{footnote}{-1}%
	\endgroup
}
\begin{document}
\allsectionsfont{\sffamily}

%
%

\title{Three-slab model for the dielectric permittivity of a lipid bilayer}

\shorttitle{Three-slab model for the dielectric permittivity of a lipid bilayer}

\author[1]{M.\ M.\ B.\ Sheraj$\,^{\dag,}$}
\author[2]{Amaresh Sahu$\,^{\ddag,}$}

\shortauthor{M.M.B.\ Sheraj and A.\ Sahu}

\affil[1]{Department of Physics, University of Texas, Austin TX 78712, USA\vspace{1pt}}
\affil[2]{McKetta Department of Chemical Engineering, University of Texas, Austin TX 78712, USA}

\date{29 July 2026}

%
%

\twocolumn[
	\begin{@twocolumnfalse}
		\maketitle
		\begin{abstract}
			\noindent\textsf{\textbf{Abstract.}}
A model for the tensorial dielectric permittivity of phospholipid membranes is presented here.
The four-nanometer-thick membrane is treated as a composite made up of three dielectric slabs: one for each of the two phospholipid head-group regions, and one for the entire domain spanned by the lipid tails.
Equal and opposite bound surface charge densities surround each head-group slab, and account for the membrane dipole potential.
Three-slab model parameters are obtained from molecular dynamics simulations, and capture both the zero-field electric potential and the membrane response to applied electric fields.
The tail region is well-approximated as having vacuum permittivity, while the head-group region is highly anisotropic due to the configurations of molecular dipoles.
For the bilayers studied, the out-of-plane permittivity of the head-group region is 10--15 times that of the vacuum, while the in-plane permittivity is an order of magnitude larger.
Membrane responses to applied electric fields up to 30 millivolts per nanometer are found to be in the linear regime.
The model overcomes a fundamental limitation of microscopic theories---where the out-of-plane permittivity lacks a meaningful continuum interpretation in the head-group region due to large gradients in the local electric field---by averaging over slab widths, thereby introducing new length scales.
Our approach can be extended to characterize interfacial systems with similar microscopic permittivities.
		\end{abstract}
		\vskip 1.6em
	\end{@twocolumnfalse}
]
\thispagestyle{empty}

\email{$^\dag \,$\href{mailto:sheraj.physics@utexas.edu}{\texttt{sheraj.physics@utexas.edu}}}
\email{$^\ddag \,$\href{mailto:asahu@che.utexas.edu}{\texttt{asahu@che.utexas.edu}}}

\normalsize
\vspace{-15pt}

%
%

\begin{figure}[!b]
	\centering\vspace{-15pt}
	\includegraphics[width=\columnwidth]{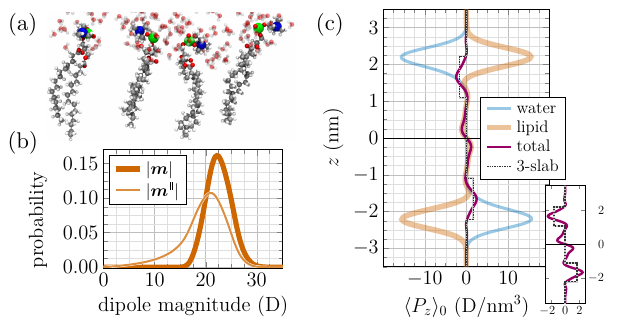}\vspace{-4pt}
	\caption{%
        DPPC phospholipid bilayer under zero applied field, from MD simulations.
        (a) Representative structure of phospholipid molecules in the upper leaflet, along with nearby water molecules.
        Each molecule has an electric dipole moment, which is separated into in-plane and out-of-plane components as
        $ \bmm = \bmmpar + \mperp \mk \bmez $.
        (b) Probability distribution of $ \lvert \bmm \rvert $ and $ \lvert \bmmpar \rvert $ for phospholipid molecules.
        (c) Average out-of-plane polarization density $ \langle \Pz \rangle_0 $ of water, lipids, and the combined system.
        While phospholipid dipoles in the top (resp.\ bottom) leaflet point up (resp.\ down), intercalated water molecules are oriented in the opposite direction and dominate the overall polarization density.
        Inset: $ \langle \Pz \rangle_0 $ for the system and three-slab model.
	}
	\label{fig_polarization}
\end{figure}

\noindent
\textsf{\textbf{Introduction.}}---%
Cells and their internal organelles are surrounded by biological membranes: two-dimensional structures that partition the various electrolytic intracellular and extracellular compartments.
Phospholipid molecules, which consist of a hydrophilic head group bound to two hydrophobic tails, are a major component of these membranes and lead to their bilayer structure.
Importantly, the oily interior of the bilayer prevents the passage of charged species, and so electric potential differences are readily established across the membrane.
A wide array of biological phenomena rely on such potential differences \cite{alberts-mboc-2022, hille-ionchannels-2001, malmivuo-bioem-1995}, and the dielectric permittivity of the membrane is central to their description.
Prior experimental \cite{okhi-enzym-2003, merla-bioem-2009, gramse-bpj-2013} and theoretical \cite{huang-bpj-1977, nymeyer-bpj-2008, edmiston-2011, steigmann-mmcs-2016, omar-pre-2024-i, omar-pre-2025-ii, omar-pre-2025-iii, yu-pre-2025, row-prr-2025, farhadi-pre-2025, fernandes-prr-2025} studies largely treat phospholipid membranes as a single slab with scalar permittivity between two and five times the vacuum permittivity $ \epsz $.
Such an idealization, however, fails to account for three features of the bilayer structure.
First, the head group is significantly more polarizable than the lipid tails \cite{wiener-bpj-1992, stern-jcp-2003}.
Second, the dipole moment of zwitterionic lipids is predominantly tangent to the membrane surface \cite{wiener-bpj-1992}, and so the head-group region exhibits an anisotropic polarization density \cite{raudino-bpj-1986, stern-jcp-2003}.
Finally, even when no external potential is applied, the out-of-plane polarization density is nonzero (see Fig.\ \figpart{fig_polarization}{c}).
As these observations have not yet been incorporated into membrane theories, their effect on biological phenomena remains unexplored.

\begin{figure}[!b]
	\centering\vspace{-9pt}
	\includegraphics[width=\columnwidth]{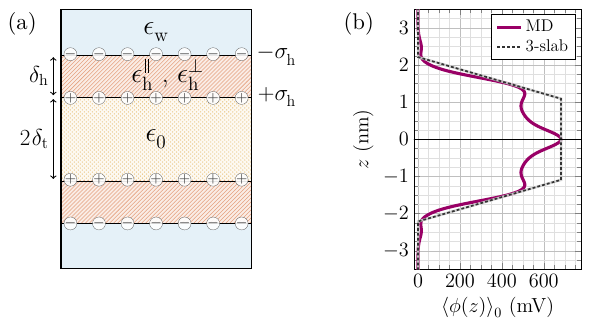}\vspace{-8pt}
	\caption{%
        Three-slab model parametrized with results from molecular simulations.
        (a) Model parameters and geometry, as described in the text.
        (b) Zero-field electric potential of a DPPC membrane as a function of distance $ z $ from the bilayer midplane.
        Three-slab model parameters are chosen to match the MD dipole potential,
        $
        \langle \phi (z = 0) \rangle^{}_0
        - \langle \phi (z = \ell/2) \rangle^{}_0 $,
        where $ \ell $ is the height of the MD unit cell.
	}
	\label{fig_three_slab}
\end{figure}

In this Communication, we develop a model for the permittivity of a zwitterionic phospholipid bilayer surrounded by water.
Microscopic features of membrane electrostatics are obtained from all-atom molecular dynamics (MD) simulations via established statistical mechanical methods \cite{stern-jcp-2003}.
The local, position-dependent permittivity tensor is calculated in this manner.
However, a local permittivity is difficult to interpret macroscopically when there are gradients in the intrinsic electric field over atomic length scales, as is the case here.
This issue is resolved by coarse-graining fine-scale details and modeling the membrane as a composite material made up of three slabs: one for the entire hydrophobic interior, and one each for the head-group region in the top and bottom leaflets (see Fig.\ \figpart{fig_three_slab}{a}).
The intrinsic polarization density from Fig.\ \figpart{fig_polarization}{c} is accounted for with surface bound charge densities at slab boundaries, and the head-group regions are characterized by a tensorial dielectric permittivity.
The three-slab model captures the overall membrane response to in-plane and out-of-plane electric fields, and we additionally determine the field beyond which the membrane response is nonlinear.
We validate the model against MD simulations of dipalmitoylphosphatidylcholine (DPPC) bilayers in the fluid phase, with similar results for dioleoylphosphatidylcholine (DOPC) membranes presented in the Supplementary Material (SM) \cite{supplemental}.

%
%

\smallskip\noindent
\textsf{\textbf{Statistical mechanics and electrostatics.}}---%
Following the seminal work of \textsc{H.A.\ Stern} and \textsc{S.E.\ Feller} \cite{stern-jcp-2003}, let $ \bmGamma $ denote the microstate of the membrane and water system---consisting of all atomic positions and momenta.
With the Hamiltonian $ \mchz (\bmGamma) $ and polarization density $ \bmP (\bmr, \bmGamma) $ at the position
$ \bmr \in \mathbb{R}^3 $,
the canonical-ensemble--averaged polarization density $ \langle \bmP (\bmr) \rangle^{}_0 $ is given by
\begin{equation} \label{eq_polarization}
\big\langle \bmP (\bmr) \big\rangle^{}_0
\, = \,
\int \! \td \bmGamma \Big( e^{- \beta \mchz} \, \bmP (\bmr, \bmGamma) \Big)
\bigg/
\int \! \td \bmGamma \, e^{- \beta \mchz}
~.
\end{equation}
Here and from now on, a subscript `0' indicates no external electric field is applied, and the inverse temperature
$ \beta := (\kBT)^{-1} $
for Boltzmann constant $ \kB $ and absolute temperature $ \vartheta $.
We take the membrane to be fluctuating about the $ x $--$ y $ plane, along which the system is translationally invariant; ensemble-averaged quantities then depend only on the distance $ z $ from the bilayer midplane.
When a constant electric field $ \bmEz $ is applied, the Hamiltonian
$ \mch (\bmGamma, \bmEz) := \mchz (\bmGamma) - \bmM (\bmGamma) \bmcdot \bmEz $,
where
$ \bmM (\bmGamma) := \int \! \td \bmr \mk \bmP (\bmr, \bmGamma) $
is the total dipole moment of the system.
If the field $ \bmEz $ is small, the change in average polarization density
$
\langle \Delta \bmP (z) \rangle_{\mkn \bmEz}
:= \langle \bmP (z) \rangle_{\mkn \bmEz}
- \langle \bmP (z) \rangle^{}_0
$
can be expressed as
\begin{equation} \label{eq_lin_resp}
\big\langle \Delta \bmP (z) \big\rangle_{\mkn \bmEz}
\, = \  \bmpsi (z) \, \bmEz
~,
\end{equation}
with
$
\bmpsi (z)
:= \beta \mk (
\langle \bmP (z) \otimes \bmM \rangle_0
- \langle \bmP (z) \rangle_0 \otimes \langle \bmM \rangle_0
)
$ \cite{stern-jcp-2003, limmer}.
Since lipid membranes in the fluid phase are additionally isotropic in-plane,
$ \langle \Px \rangle_0 = \langle \Py \rangle_0 = 0 $
and
$ \bmpsi $ is diagonal with two independent components---which we denote
$ \psipar(z) := \psi_{x x} (z) = \psi_{y y} (z) $
and
$ \psiperp(z) := \psi_{z z} (z) $.
From the definition of $ \bmpsi $ below Eq.\ \eqref{eq_lin_resp}, $ \psipar $ and $ \psiperp $ can be determined from MD simulations.
Our task now is to connect $ \psipar $ and $ \psiperp $ to the membrane dielectric permittivity tensor $ \bmeps $---which is also diagonal, with the two independent components $ \epspar (z) $ and $ \epsperp (z) $.

By definition, for systems with an intrinsic polarization density, the permittivity relates the changes in average polarization density and average electric field as
\begin{equation} \label{eq_permittivity_def}
\big\langle \Delta \bmP (z) \big\rangle_{\mkn \bmEz}
\, = \, \big( \bmeps (z) \mk - \mk \epsz \mk \bmI \big) \mk \big\langle \Delta \bmE (z) \big\rangle_{\mkn \bmEz}
~,
\end{equation}
where $ \bmI $ is the identity tensor.
In addition, for MD simulations of a system with volume $ V $, a constant electric field $ \bmEz $ entering the Hamiltonian causes a change in the average field $ \langle \Delta \bmE \rangle_{\mkn \bmEz} $ according to \cite{stern-jcp-2003}
\begin{equation} \label{eq_DeltaE}
\big\langle \Delta \bmE (z) \big\rangle_{\mkn \bmEz}
\, = \, \bmEz
\, + \, \dfrac{\bmez}{\epsz} \bigg(
\dfrac{\langle \Delta M_z \rangle_{\mkn \bmEz}}{V}
\, - \, \big\langle \Delta \Pz (z) \big\rangle_{\mkn \bmEz}
\bigg)
~,
\end{equation}
where $ \bmez $ is the unit vector along the $ z $-axis and $ \langle \Delta M_z \rangle_{\mkn \bmEz} $ is the change in the $ z $-component of the total dipole moment of the system \cite{conducting-bcs}.
Combining Eqs.\ \eqref{eq_lin_resp}--\eqref{eq_DeltaE} yields a relation between $ \bmpsi $ and $ \bmeps $---though it is nontrivial due to the local and average polarization densities entering the right-hand side of Eq.\ \eqref{eq_DeltaE}.
We now separately consider the membrane response to in-plane and out-of-plane electric fields to determine $ \epspar (z) $ and $ \epsperp (z) $.

\begin{figure}[!b]
	\centering 
	\includegraphics[width=\columnwidth]{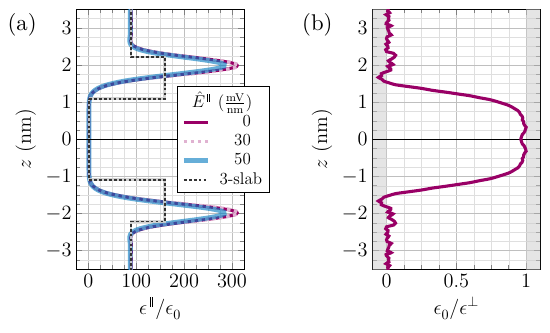}%
    \vspace{-4pt}
	\caption{%
        Permittivity of a DPPC membrane.
        (a) The in-plane permittivity $ \epspar (z) $ is calculated from MD simulations via the linear response and fluctuation formulas in Eq.~\eqref{eq_epspar}.
        The two approaches agree for electric fields up to 30 mV$/$nm, beyond which the system response is nonlinear (see the SM \cite{supplemental}).
        The in-plane permittivity of the three-slab model (thin dashed line) captures essential features of the MD result according to Eq.\ \eqref{eq_in_plane}.
        (b) Reciprocal of the out-of-plane permittivity $ \epsperp (z) $, as calculated from Eq.\ \eqref{eq_epsperp_fluct}.
        In the head-group region, $ \epsperp $ takes macroscopically unphysical values (grey shading), for which $ 1 / \epsperp $ crosses zero.
	}
	\label{fig_eps}
\end{figure}

To calculate $ \epspar $, let $ \bmEz $ be parallel to the membrane; we choose
$ \bmEz = \Ez \mk \bmex $
without loss of generality.
According to Eqs.\ \eqref{eq_lin_resp}--\eqref{eq_DeltaE},
$
\langle \Delta \Px \rangle_{\mkn \Ez}
= \langle \Px \rangle_{\mkn \Ez}
= \psipar \mk \Ez
= ( \epspar - \epsz ) \mk \Ez
$,
from which we obtain the two relations \cite{stern-jcp-2003}
\begin{equation} \label{eq_epspar}
\epspar (z)
\mk = \mk \epsz
\mk + \mk \dfrac{\langle \Px (z) \rangle_{\! \Ez}}{\Ez}
~\quad~
\text{and}
~\quad~
\epspar (z)
\mk = \mk \epsz
\mk + \mk \psipar (z)
\, .
\end{equation}
The first expression in Eq.\ \eqref{eq_epspar} determines $ \epspar $ via linear response: an in-plane electric field is applied in simulation, and the polarization density is measured.
In doing so, care must be taken to ensure the applied field is small, such that the response is indeed linear.
In contrast, the second expression in Eq.\ \eqref{eq_epspar} calculates $ \epspar $ from equilibrium fluctuations with no applied field.
Figure \figpart{fig_eps}{a} confirms the two methods agree for small $ \Ez $, and reveals the membrane response is linear for in-plane fields up to 30 mV$/$nm.
The in-plane permittivity in the head-group region is significantly larger than that in the lipid tails \cite{stern-jcp-2003, loche-jpcb-2020}, which motivates subsequent development of the three-slab model.

The out-of-plane permittivity is calculated via a similar procedure.
For an external electric field
$ \bmEz = \Ez \mk \bmez $,
we combine Eqs.\ \eqref{eq_lin_resp} and \eqref{eq_permittivity_def} to obtain
$
\langle \Delta \Pz \rangle_{\mkn \Ez}
= \psiperp \mk \Ez
= (\epsperp - \epsz) \mk \langle \Delta E_z \rangle_{\mkn \Ez}
$.
Substituting Eq.\ \eqref{eq_DeltaE} then yields \cite{stern-jcp-2003}
\begin{gather}
\epsperp (z)
\, = \, \epsz
\, + \, \dfrac{
\epsz \, \langle \Delta \Pz (z) \rangle_{\mkn \Ez}
}{
\epsz \mk \Ez
\, - \, \langle \Delta \Pz (z) \rangle_{\mkn \Ez}
\, + \, \langle \Delta M_z \rangle_{\mkn \Ez} / V
}
\label{eq_epsperp_lin}
\intertext{and}
\epsperp (z)
\, = \, \epsz
\, + \, \dfrac{
\epsz \, \psiperp (z)
}{
\epsz
\, - \, \psiperp (z)
\, + \, \psiperpbar
}
~,
\label{eq_epsperp_fluct}
\end{gather}
where
$ \psiperpbar \mkn := \int \mkn \td \bmr \mk \psiperp (z) / V $
is the average value of $ \psiperp (z) $.
Equations \eqref{eq_epsperp_lin} and \eqref{eq_epsperp_fluct} are the linear response and fluctuation formulas for $ \epsperp $, respectively.
However, in calculating the out-of-plane permittivity from MD data via Eq.\ \eqref{eq_epsperp_fluct}, we find
$ 1 / \epsperp $ crosses zero in the head-group region (see Fig.\ \figpart{fig_eps}{b}).
Consequently, $ \epsperp $ is unbounded and takes negative values---which, from a macroscopic perspective, is unphysical.
We confirm via extensive MD simulations \cite{supplemental} that such features are not statistical artifacts, and are in fact experimentally relevant \cite{fellows-jpcc-2024}; similar observations were reported in simulations of water near surfaces \cite{nienhuis-jcp-1971, ballenegger-jcp-2005, bonthuis-prl-2011, bonthuis-l-2012, loche-prl-2019}, including lipid bilayers \cite{loche-jpcb-2020}.

While $ \epsperp (z) $ is meaningful as a local quantity, it cannot easily be related to the macroscopic response of the membrane.
To understand why, consider the continuum perspective, where the permittivity is a macroscopic quantity.
In such a description, a spatially-dependent permittivity relates, at each material point, the mean polarization density and electric fields---averaged over a coarse-graining volume in which the fields are assumed to not vary appreciably.
In the present scenario, however, there is no such separation in length scales because $ \langle E_z (z) \rangle $ varies considerably over atomic distances \cite{supplemental}.
Consequently, there is not a straightforward relationship between $ \epsperp (z) $ and the macroscopic membrane response to an out-of-plane electric field.

%
%

\smallskip\noindent
\textsf{\textbf{Three-slab model.}}---%
The coarse-grained behavior of a lipid membrane is not readily obtained from the local permittivity tensor $ \bmeps (z) $ for the aforementioned reasons.
However, Figs.\ \ref{fig_polarization}--\ref{fig_eps} show bilayer properties are also far from uniform across the membrane thickness.
A fundamental question thus remains:
How can the macroscopic response of a phospholipid bilayer to external electric fields be quantified?
We address this question by recognizing membrane features are qualitatively different within the head-group and tail regions, and thus modeling the membrane as a composite material made up of three uniform dielectric slabs (see Fig.\ \figpart{fig_three_slab}{a}).
There is one slab of thickness $ \deltah $ for each of the two head-group regions, while a single slab of thickness $ 2 \mk \deltat $ represents the entire tail region.
With our three-slab construction, microscopic features are averaged over lengths larger than the atomic distances over which $ \langle E_z (z) \rangle $ varies, and smaller than the bilayer thickness---such that the out-of-plane heterogeneity of the membrane is retained.
We note multi-slab models were previously used to describe the permittivity of water near solid surfaces \cite{schlaich-prl-2016, loche-jpcb-2020, jalali-pre-2020,stark-cpr-2026}.

In the absence of an external electric field, a material comprised of three dielectric slabs is unpolarized and at a constant electric potential throughout.
Both features are inconsistent with the MD data in Figs.\ \figpart{fig_polarization}{c} and \figpart{fig_three_slab}{b}.
We thus introduce an intrinsic, zero-field polarization density into the three-slab model.
For simplicity, the polarization density is constant and limited to the head-group slab.
It is captured by equal and opposite bound surface charge densities of magnitude $ \sigmah $ on the faces of the head-group slabs, as shown in Fig.\ \figpart{fig_three_slab}{a}.
We take the permittivity of the tail slab to be $ \epsz \mk \bmI $, as the MD results in Fig.\ \ref{fig_eps} show $ \epspar (z) / \epsz $ and $ \epsperp (z) / \epsz $ are just above unity in the tail region.
In contrast, the head-group slabs in our model have an anisotropic permittivity, with in-plane and out-of-plane components respectively denoted $ \epshpar $ and $ \epshperp $.
The three-slab model accordingly captures key electrostatic features of the membrane.

%
%

\smallskip\noindent
\textsf{\textbf{Determining model parameters.}}---%
Our model is defined by the parameters $ \deltat $, $ \deltah $, $ \epshpar $, $ \epshperp $, and $ \sigmah $.
We assume the membrane thickness
$ \deltam := 2 (\deltat + \deltah) $
is known a priori, either from standard methods \cite{egberts-ebj-1994} or as described below.
In what follows, the independent parameters $ \deltah $, $ \epshpar $, $ \epshperp $, and $ \sigmah $ are solved for by equating relevant features from MD simulations with their three-slab counterparts.
We thus obtain the three-slab parameters of DPPC and DOPC bilayers, as presented in Table \ref{tab_params}.

We first consider the membrane at zero external field, for which both $ \langle \Pz (z) \rangle_0 $ and $ \langle \phi (z) \rangle_0 $ are nontrivial.
Since $ \langle \Pz \rangle_0 $ is an odd function of $ z $, we choose to capture its integral from the bilayer midplane
(at $ z = 0 $)
to the top of the MD unit cell
(at $ z = \ell / 2 $).
As there is no free charge in the system,
$ \langle \phi''(z) \rangle_0 = \langle \Pz {\!}'(z) \rangle_0 / \epsz $
and \cite{supplemental}
\begin{equation} \label{eq_dipole_potential}
\big\langle \phi (0) \big\rangle_0
\, - \, \big\langle \phi (\ell / 2) \big\rangle_0
\, = \, \dfrac{\sigmah \mk \deltah}{\epsz}
~.
\end{equation}
The left-hand side of Eq.\ \eqref{eq_dipole_potential} is the so-called \textit{dipole potential,} and is calculated from MD simulations, while the right-hand side is in terms of three-slab parameters.
In this way, the three-slab model captures key features of the zero-field membrane, as shown in Figs.\ \figpart{fig_polarization}{c} and~\figpart{fig_three_slab}{b}.

\begin{figure}[!b]
	\centering\vspace{-10pt}
	\includegraphics[width=\columnwidth]{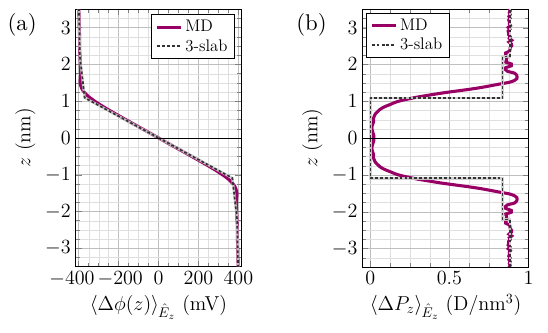}%
    \vspace{-8pt}
	\caption{%
        Change in electric potential (a) and out-of-plane polarization density (b) of a DPPC bilayer due to an electric field
        $ \bmEz = \Ezz \bmez $.
        The three-slab parameters (see Table \ref{tab_params}) are obtained with
        $ \Ezz = 20 $ mV$/$nm,
        and well-approximate the results here, for which
        $ \Ezz = 70 $ mV$/$nm.
        (a) The model captures the MD response, with
        $ \langle \Delta E_z \rangle := - \langle \Delta \phi' (z) \rangle $
        largely confined to the tail region.
        (b) The change in polarization is predominantly in the head-group regions and surrounding water, and is reasonably described by the model.
	}
	\label{fig_Delta}
\end{figure}

Next, the membrane response to electric fields is investigated.
For an in-plane field with magnitude $ \Ezpar $, we equate the integral of the in-plane polarization density---or, equivalently, the in-plane permittivity---in the three-slab and MD scenarios.
Introducing
$ \deltaw := \ell/2 - \deltah - \deltat $
as the height of the water layer above (or below) the membrane and $ \epsw $ as the water permittivity, we obtain \cite{supplemental}
\begin{equation} \label{eq_in_plane}
\int_0^{\ell/2} \! \epspar(z) ~ \td z
\, = \, \deltat \mk \epsz
\, + \, \deltah \mk \epshpar
\, + \, \deltaw \mk \epsw
~.
\end{equation}
Figure \figpart{fig_eps}{a} compares $ \epspar $ with its three-slab counterpart.
If instead the field is orthogonal to the membrane, for which
$ \bmEz = \Ezz \bmez $,
the out-of-plane polarization density is altered from the zero-field scenario.
Since the integral of the latter is accounted for in Eq.\ \eqref{eq_dipole_potential}, we now equate the integral of the change in polarization density in MD simulations and the three-slab model, which yields \cite{supplemental}
\begin{align} \label{eq_DeltaPz}
\int\displaylimits_0^{\ell/2} \dfrac{
\langle \Delta \Pz (z) \rangle_{\! \Ezz} \!
}{
\Ezz \mk ( \ell / 2 )
} ~ \td z
\, = \, \dfrac{
\deltah \mk (\epshperp - \epsz)
+ \deltaw \mk (\epsw - \epsz) \mk \epshperp \mkn / \epsw
}{
\epshperp (\deltat / \epsz \, + \, \deltah / \epshperp \, + \, \deltaw / \epsw)
}
\,.
\notag
\\[-10pt]
\end{align}

With Eqs.\ \eqref{eq_dipole_potential}--\eqref{eq_DeltaPz}, we account for the membrane's zero-field polarization, as well as its change in polarization due to parallel and perpendicular electric fields.
One additional relation is needed to solve for the three-slab parameters.
We choose to compare the first moment of $ \langle \Delta \Pz (z) \rangle_{\! \Ezz} $ in MD simulations with its three-slab counterpart, for which \cite{supplemental}
\begin{align} \label{eq_zDeltaPz}
&\int_0^{\ell/2} \dfrac{
z \mk \langle \Delta \Pz (z) \rangle_{\! \Ezz} \!
}{
\Ezz \mk ( \ell / 2 )
} ~ \td z
\\[4pt]
&= \, \dfrac{
(\deltah^{\, 2} + 2 \mk \deltat \mk \deltah) \mk (\epshperp - \epsz)
+ \tfrac{1}{4} \mk (\ell^2 - \deltam^2) \mk (\epsw - \epsz) \mk \epshperp \mkn / \epsw
}{
2 \mk \epshperp (\deltat / \epsz \, + \, \deltah / \epshperp \, + \, \deltaw / \epsw)
}
\, .
\notag
\end{align}
In a multipole expansion of the electric potential resulting from the upper bilayer leaflet
($ 0 \le z \le \ell / 2 $),
Eq.\ \eqref{eq_DeltaPz} captures the dipole moment resulting from $ \Ezz $, while Eq.\ \eqref{eq_zDeltaPz} captures the induced quadrupole moment.
We thus have four equations (\ref{eq_dipole_potential}--\ref{eq_zDeltaPz}) involving the four independent parameters of the three-slab model.
In the SM \cite{supplemental}, we examine how three-slab parameters are modified when Eq.\ \eqref{eq_zDeltaPz} is replaced by a related constraint.

Recall that at the outset, the membrane thickness $ \deltam $ was assumed to be known.
For a DPPC bilayer, standard definitions based on atomic positions \cite{egberts-ebj-1994} yield a thickness of 3.9--4.2 nm.
In contrast, the electrostatic influence of the membrane suggests a thickness closer to 5 nm, as seen in Figs.\ \ref{fig_polarization}--\ref{fig_eps}.
In an effort to bridge these perspectives, we define the membrane--water interface as where
$ \td^2 \epspar (z) / \td z^2 = 0 $
in the MD data---for which
$ \deltam \approx 4.4 $ nm.
In the SM \cite{supplemental}, we verify the three-slab results remain valid for different definitions of $ \deltam $.

\begin{table}[!t]
	\centering
    \caption{%
        Three-slab model parameters for DPPC and DOPC bilayers, calculated from Eqs.\ \eqref{eq_dipole_potential}--\eqref{eq_zDeltaPz} with
        $ \Ezz = 20 $ mV$/$nm.
        The bilayers are respectively at 323 K and 298 K, and so are in the fluid phase; see the SM \cite{supplemental} for simulation parameters.%
    }%
    \vspace{-7pt}
    \setlength{\tabcolsep}{8pt}
    \renewcommand{\arraystretch}{1.05}
    \begin{tabular}{c | c c c c c}
        \hline
        \hline
        \\[-12pt]
        ~
        &
        $ \deltat $
        &
        $ \deltah $
        &
        $ \epshpar $
        &
        $ \epshperp $
        &
        $ \sigmah $
        \\[1pt]
        ~
        &
        \small(nm)
        &
        \small(nm)
        &
        \small($ \epsz $)
        &
        \small($ \epsz $)
        &
        \small($ \text{e} / \text{nm}^2 $)
        \\
        \hline
        \\[-10pt]
        DPPC &
        1.1 &
        1.1 &
        160 &
        16 &
        0.034
        \\[1pt]
        DOPC &
        1.0 &
        1.1 &
        170 &
        9.2 &
        0.029 
        \\[1pt]
    \hline
    \hline
    \end{tabular}%
    \vspace{-7pt}
    \label{tab_params}
\end{table}

%
%

\smallskip\noindent
\textsf{\textbf{Results and discussion.}}---%
Three-slab model parameters for DPPC and DOPC bilayers are presented in Table \ref{tab_params}.
Within each leaflet, the widths of the head-group and tail slabs are comparable.
Moreover,
$ \epshperp < \epsw < \epshpar $
and the in-plane and out-of-plane permittivities differ by an order of magnitude in the head-group slabs.
To relate three-slab parameters with experimental measurements, we calculate the membrane capacitance per unit area as
$
    \Cm
    := (2 \mk \deltat / \epsz
    + 2 \mk \deltah / \epshperp)^{-1}
    \approx 0.4 \text{ \textmu F/cm}^2
$
for both DPPC and DOPC---in reasonable agreement with experiments \cite{okhi-enzym-2003, naumowicz-electrochim-2006, merla-bioem-2009, gramse-bpj-2013, muzio-langmuir-2020}.
One can then define an effective out-of-plane membrane permittivity as
$ \epsmperp := \deltam \mk \Cm \approx 2 \mk \epsz $,
which is again consistent with experimental values \cite{okhi-enzym-2003, merla-bioem-2009, gramse-bpj-2013}.
However, we reiterate that the membrane should not be described as a single dielectric slab of constant, scalar permittivity.

To test the robustness of the model, its features are compared to MD results in Figs.\ \ref{fig_polarization}--\ref{fig_Delta} (for DPPC) and the SM \cite{supplemental} (for DOPC).
As expected, the zero-field MD results in Figs.\ \figpart{fig_polarization}{c} and \figpart{fig_three_slab}{b} are only approximately captured by our coarse-grained model---in which properties are uniform across each slab.
Importantly, despite $ \epsperp (z) $ diverging at microscopic scales, the three-slab model captures the overall change in potential and out-of-plane polarization density due to an out-of-plane electric field (see Fig.\ \ref{fig_Delta}).
Here the three-slab parameters are obtained via Eqs.\ \eqref{eq_dipole_potential}--\eqref{eq_zDeltaPz} when
$ \Ezz = 20 $ mV$/$nm
in MD simulations, while the data in Fig.\ \ref{fig_Delta} is generated with a field of magnitude 70 mV$/$nm.
We thus confirm that for out-of-plane fields, each slab in our model can be treated as a linear dielectric material at least until
$ \Ezz = 70 $ mV$/$nm.
In contrast, the in-plane MD analysis shown in Fig.\ \figpart{fig_eps}{a} reveals the in-plane response is linear for fields with magnitude 30 mV$/$nm or less.
As the three-slab model incorporates data from $ \epspar (z) $ via Eq.\ \eqref{eq_in_plane}, it is similarly limited for in-plane fields.

%
%

\smallskip\noindent
\textsf{\textbf{Conclusions.}}---%
In this Communication, we developed a model for the dielectric permittivity of a lipid membrane, in which a bilayer is treated as three dielectric slabs.
The central slab represents the oily membrane interior, and has vacuum permittivity.
Outer slabs model the phospholipid heads, and have anisotropic permittivities and intrinsic polarization densities.
We calculated a physically meaningful macroscopic permittivity $ \epshperp $ in the head-group region, despite the divergence of $ \epsperp (z) $ there, by averaging microscopic features over slab thicknesses.
With all-atom MD simulations, we verified that the three-slab model captures zero-field membrane properties and the average bilayer response to electric fields.
The three-slab membrane capacitance was also shown to be consistent with experiments.

Since our model involves only a few parameters, it can be incorporated into continuum theories of membrane electromechanics \cite{edmiston-2011, steigmann-mmcs-2016, omar-pre-2024-i, omar-pre-2025-ii, omar-pre-2025-iii, yu-pre-2025}.
Doing so could shed light on bilayer flexoelectricity \cite{petrov-aca-2006, harland-pre-2010, thomas-epl-2021, torbati-rmp-2022}, as well as electric field-induced vesicle deformations \cite{dimova-sm-2007, dimova-sm-2009} and phase transition modifications \cite{thomas-sm-2022}.
Before investigating such complex phenomena, it would be natural to first extend the three-slab model to biologically-relevant systems with ions in the surrounding fluid and multiple phospholipid species, some of which carry a net charge.
One could also examine whether the local permittivity of water near solid surfaces \cite{nienhuis-jcp-1971, ballenegger-jcp-2005, bonthuis-prl-2011, bonthuis-l-2012, loche-prl-2019, loche-jpcb-2020} can similarly be resolved with a slab-like approach.

%
%




\small
\smallskip\balance

%
%

\vspace{15pt}

\noindent\textbf{\textsf{Acknowledgments.}}
It is a pleasure to thank
Prof.\ \href{https://mandadapu-group.github.io/}{Kranthi Mandadapu}
for insightful conversations on membrane electrostatics and comments on the manuscript, as well as
Prof.\ \href{https://physiology.med.cornell.edu/people/emre-aksay-ph-d/}{Emre Aksay}
for many discussions on electrically excitable membranes.
We thank
Profs.\ \href{https://chemistry.berkeley.edu/people/david-limmer}{David Limmer}
and
\href{https://sites.utexas.edu/ganesan/}{Venkat Ganesan}
for helpful interactions, including bringing several references to our attention.

This work was supported by the \href{https://welch1.org/}{Welch
Foundation} via Grant No.\ F-2208.
We are grateful to the \href{https://tacc.utexas.edu}{Texas Advanced Computing Cluster}, where all simulations were carried out.

~

%
%

\small%
\subsection*{References}
\bibliographystyle{bibStyle}
\bibliography{refs}

\begin{thebibliography}{10}
\expandafter\ifx\csname url\endcsname\relax
  \def\url#1{\texttt{#1}}\fi
\expandafter\ifx\csname urlprefix\endcsname\relax\def\urlprefix{URL }\fi
\providecommand{\bibinfo}[2]{#2}
\providecommand{\eprint}[2][]{\texttt{\href{https://arxiv.org/abs/#2}{#2}}}

\bibitem{alberts-mboc-2022}
\bibinfo{author}{B. Alberts} \emph{et~al.}
\newblock
  \href{https://wwnorton.com/books/9780393884821}{\emph{\bibinfo{title}{{Molecular
  Biology of the Cell}}}} (\bibinfo{publisher}{W.W.\ Norton, Inc.},
  \bibinfo{address}{New York}, \bibinfo{year}{2022}), \bibinfo{edition}{7th}
  edn.

\bibitem{hille-ionchannels-2001}
\bibinfo{author}{B. Hille}.
\newblock
  \href{https://digitalcommons.rockefeller.edu/ru-authors/92/}{\emph{\bibinfo{title}{{Ion
  Channels of Excitable Membranes}}}} (\bibinfo{publisher}{Sinauer Associates,
  Inc.}, \bibinfo{address}{Sunderland}, \bibinfo{year}{2001}),
  \bibinfo{edition}{3rd} edn.

\bibitem{malmivuo-bioem-1995}
\bibinfo{author}{J. Malmivuo} \& \bibinfo{author}{R. Plonsey}.
\newblock
  \href{https://doi.org/10.1093/acprof:oso/9780195058239.001.0001}{\emph{\bibinfo{title}{Bioelectromagnetism:
  Principles and Applications of Bioelectric and Biomagnetic Fields}}}
  (\bibinfo{publisher}{Oxford University Press}, \bibinfo{address}{New York},
  \bibinfo{year}{1995}).

\bibitem{okhi-enzym-2003}
\bibinfo{author}{S. Ohki} \& \bibinfo{author}{K. Arnold}.
\newblock
  \href{https://doi.org/10.1016/S0076-6879(03)67016-3}{\bibinfo{title}{Determination
  of liposome surface dielectric constant and hydrophobicity}}.
\newblock In: \emph{\bibinfo{booktitle}{Liposomes, Part A}}, vol.
  \bibinfo{volume}{367} of \emph{\bibinfo{series}{Methods Enzymol.}},
  \bibinfo{pages}{253--272} (\bibinfo{publisher}{Academic Press},
  \bibinfo{year}{2003}).

\bibitem{merla-bioem-2009}
\bibinfo{author}{C. Merla}, \bibinfo{author}{M. Liberti}, \bibinfo{author}{F.
  Apollonio} \& \bibinfo{author}{G. d’Inzeo}.
\newblock
  \emph{\bibinfo{journal}{\href{https://doi.org/10.1002/bem.20476}{Bioelectromagnetics}}}
  \textbf{\bibinfo{volume}{30}}, \bibinfo{pages}{286--298}
  (\bibinfo{year}{2009}).

\bibitem{gramse-bpj-2013}
\bibinfo{author}{G. Gramse}, \bibinfo{author}{A. Dols-Perez},
  \bibinfo{author}{M. Edwards}, \bibinfo{author}{L. Fumagalli} \&
  \bibinfo{author}{G. Gomila}.
\newblock
  \emph{\bibinfo{journal}{\href{https://doi.org/10.1016/j.bpj.2013.02.011}{Biophys.\
  J.}}} \textbf{\bibinfo{volume}{104}}, \bibinfo{pages}{1257--1262}
  (\bibinfo{year}{2013}).

\bibitem{huang-bpj-1977}
\bibinfo{author}{W. Huang} \& \bibinfo{author}{D.~G. Levitt}.
\newblock
  \emph{\bibinfo{journal}{\href{https://doi.org/10.1016/S0006-3495(77)85630-0}{Biophys.\
  J.}}} \textbf{\bibinfo{volume}{17}}, \bibinfo{pages}{111--128}
  (\bibinfo{year}{1977}).

\bibitem{nymeyer-bpj-2008}
\bibinfo{author}{H. Nymeyer} \& \bibinfo{author}{H.-X. Zhou}.
\newblock
  \emph{\bibinfo{journal}{\href{https://doi.org/10.1529/biophysj.107.117770}{Biophys.\
  J.}}} \textbf{\bibinfo{volume}{94}}, \bibinfo{pages}{1185--1193}
  (\bibinfo{year}{2008}).

\bibitem{edmiston-2011}
\bibinfo{author}{J. Edmiston} \& \bibinfo{author}{D.~J. Steigmann}.
\newblock
  \href{https://doi.org/10.1007/978-3-7091-0701-0}{\bibinfo{title}{Analysis of
  nonlinear electrostatic membranes}}.
\newblock In: \bibinfo{editor}{R.~W. Ogden} \& \bibinfo{editor}{D.~J.
  Steigmann} (eds.) \emph{\bibinfo{booktitle}{Mechanics and Electrodynamics of
  Magneto- and Electro-Elastic Materials}}, \bibinfo{pages}{153--180}
  (\bibinfo{publisher}{Springer}, \bibinfo{address}{New York},
  \bibinfo{year}{2011}).

\bibitem{steigmann-mmcs-2016}
\bibinfo{author}{D.~J. Steigmann} \& \bibinfo{author}{A. Agrawal}.
\newblock
  \emph{\bibinfo{journal}{\href{https://doi.org/10.2140/memocs.2016.4.31}{Math.\
  Mech.\ Complex Syst.}}} \textbf{\bibinfo{volume}{4}}, \bibinfo{pages}{31--54}
  (\bibinfo{year}{2016}).

\bibitem{omar-pre-2024-i}
\bibinfo{author}{Y.~A.~D. Omar}, \bibinfo{author}{Z.~G. Lipel} \&
  \bibinfo{author}{K.~K. Mandadapu}.
\newblock
  \emph{\bibinfo{journal}{\href{https://doi.org/10.1103/PhysRevE.109.054401}{Phys.\
  Rev.\ E}}} \textbf{\bibinfo{volume}{109}}, \bibinfo{pages}{054401}
  (\bibinfo{year}{2024}).
\newblock arXiv:\eprint{2301.09610}.

\bibitem{omar-pre-2025-ii}
\bibinfo{author}{Y.~A.~D. Omar}, \bibinfo{author}{Z.~G. Lipel} \&
  \bibinfo{author}{K.~K. Mandadapu}.
\newblock
  \emph{\bibinfo{journal}{\href{https://doi.org/10.1103/75tt-k2f5}{Phys.\ Rev.\
  E}}} \textbf{\bibinfo{volume}{112}}, \bibinfo{pages}{024406}
  (\bibinfo{year}{2025}).
\newblock arXiv:\eprint{2309.03863}.

\bibitem{omar-pre-2025-iii}
\bibinfo{author}{Y.~A.~D. Omar}, \bibinfo{author}{Z.~G. Lipel} \&
  \bibinfo{author}{K.~K. Mandadapu}.
\newblock
  \emph{\bibinfo{journal}{\href{https://doi.org/10.1103/h2hs-rg2z}{Phys.\ Rev.\
  E}}} \textbf{\bibinfo{volume}{112}}, \bibinfo{pages}{024407}
  (\bibinfo{year}{2025}).
\newblock arXiv:\eprint{2501.11612}.

\bibitem{yu-pre-2025}
\bibinfo{author}{Z. Yu}, \bibinfo{author}{S. Zhao}, \bibinfo{author}{M.~J.
  Miksis} \& \bibinfo{author}{P.~M. Vlahovska}.
\newblock
  \emph{\bibinfo{journal}{\href{https://doi.org/10.1103/91m7-tq8k}{Phys.\ Rev.\
  E}}} \textbf{\bibinfo{volume}{112}}, \bibinfo{pages}{054408}
  (\bibinfo{year}{2025}).
\newblock arXiv:\eprint{2502.12551}.

\bibitem{row-prr-2025}
\bibinfo{author}{H. Row}, \bibinfo{author}{J.~B. Fernandes},
  \bibinfo{author}{K.~K. Mandadapu} \& \bibinfo{author}{K. Shekhar}.
\newblock
  \emph{\bibinfo{journal}{\href{https://doi.org/10.1103/PhysRevResearch.7.013185}{Phys.\
  Rev.\ Res.}}} \textbf{\bibinfo{volume}{7}}, \bibinfo{pages}{013185}
  (\bibinfo{year}{2025}).
\newblock
  \bibinfo{note}{{}arXiv:\href{https://arxiv.org/abs/2407.11947}{\texttt{2407.}}
  \href{https://arxiv.org/abs/2407.11947}{\texttt{11947}}}.

\bibitem{farhadi-pre-2025}
\bibinfo{author}{J. Farhadi}, \bibinfo{author}{J.~B. Fernandes},
  \bibinfo{author}{K. Shekhar} \& \bibinfo{author}{K.~K. Mandadapu}.
\newblock
  \emph{\bibinfo{journal}{\href{https://doi.org/10.1103/PhysRevE.111.064412}{Phys.\
  Rev.\ E}}} \textbf{\bibinfo{volume}{111}}, \bibinfo{pages}{064412}
  (\bibinfo{year}{2025}).
\newblock
  \bibinfo{note}{{}arXiv:\href{https://arxiv.org/abs/2503.04677}{\texttt{2503.}}
  \href{https://arxiv.org/abs/2503.04677}{\texttt{04677}}}.

\bibitem{fernandes-prr-2025}
\bibinfo{author}{J.~B. Fernandes}, \bibinfo{author}{H. Row},
  \bibinfo{author}{K.~K. Mandadapu} \& \bibinfo{author}{K. Shekhar}.
\newblock
  \emph{\bibinfo{journal}{\href{https://doi.org/10.1103/m446-4lnj}{Phys.\ Rev.\
  Res.}}} \textbf{\bibinfo{volume}{8}}, \bibinfo{pages}{013137}
  (\bibinfo{year}{2026}).
\newblock
  \bibinfo{note}{{}arXiv:\href{https://arxiv.org/abs/2508.14001}{\texttt{2508.}}
  \href{https://arxiv.org/abs/2508.14001}{\texttt{14001}}}.

\bibitem{wiener-bpj-1992}
\bibinfo{author}{M.~C. Wiener} \& \bibinfo{author}{S.~H. White}.
\newblock
  \emph{\bibinfo{journal}{\href{https://doi.org/10.1016/S0006-3495(92)81849-0}{Biophys.\
  J.}}} \textbf{\bibinfo{volume}{61}}, \bibinfo{pages}{434--447}
  (\bibinfo{year}{1992}).

\bibitem{stern-jcp-2003}
\bibinfo{author}{H.~A. Stern} \& \bibinfo{author}{S.~E. Feller}.
\newblock \emph{\bibinfo{journal}{\href{https://doi.org/10.1063/1.1537242}{J.\
  Chem.\ Phys.}}} \textbf{\bibinfo{volume}{118}}, \bibinfo{pages}{3401--3412}
  (\bibinfo{year}{2003}).

\bibitem{raudino-bpj-1986}
\bibinfo{author}{A. Raudino} \& \bibinfo{author}{D. Mauzerall}.
\newblock
  \emph{\bibinfo{journal}{\href{https://doi.org/10.1016/S0006-3495(86)83480-4}{Biophys.\
  J.}}} \textbf{\bibinfo{volume}{50}}, \bibinfo{pages}{441--449}
  (\bibinfo{year}{1986}).

\bibitem{supplemental}
\bibinfo{title}{{See the
  \href{https://sahu-lab.github.io/pubs/three-slab-sm.pdf}{Supplementary
  Material} (SM), where relevant equations are derived, simulation protocols
  are provided, and further results are presented. The SM additionally cites
  Refs.\ \cite{kirkwood-jcp-1939, neumann-mp-1983, klauda-jpcb-2010,
  abraham-zenodo-2024, jorgensen-jcp-1983, berendsen-cpc-1995,
  feng-jctc-2023}.}}

\bibitem{limmer}
\bibinfo{author}{D.~T. Limmer}.
\newblock
  \href{https://doi.org/10.1093/oso/9780198919858.001.0001}{\emph{\bibinfo{title}{Statistical
  Mechanics and Stochastic Thermodynamics: A Textbook on Modern Approaches in
  and out of Equilibrium}}} (\bibinfo{publisher}{Oxford University Press},
  \bibinfo{year}{2024}).

\bibitem{conducting-bcs}
\bibinfo{title}{{In MD simulations, Eq.\ \eqref{eq_DeltaE} is correct only for
  conducting boundary conditions; the more general case is treated in Ref.\
  \cite{stern-jcp-2003}}}.

\bibitem{loche-jpcb-2020}
\bibinfo{author}{P. Loche}, \bibinfo{author}{C. Ayaz}, \bibinfo{author}{A.
  Wolde-Kidan}, \bibinfo{author}{A. Schlaich} \& \bibinfo{author}{R.~R. Netz}.
\newblock
  \emph{\bibinfo{journal}{\href{https://doi.org/10.1021/acs.jpcb.0c01967}{J.\
  Phys.\ Chem.\ B}}} \textbf{\bibinfo{volume}{124}},
  \bibinfo{pages}{4365--4371} (\bibinfo{year}{2020}).

\bibitem{fellows-jpcc-2024}
\bibinfo{author}{A.~P. Fellows} \emph{et~al.}
\newblock
  \emph{\bibinfo{journal}{\href{https://doi.org/10.1021/acs.jpcc.4c06650}{J.\
  Phys.\ Chem.\ C}}} \textbf{\bibinfo{volume}{128}},
  \bibinfo{pages}{20733--20750} (\bibinfo{year}{2024}).

\bibitem{nienhuis-jcp-1971}
\bibinfo{author}{G. Nienhuis} \& \bibinfo{author}{J.~M. Deutch}.
\newblock \emph{\bibinfo{journal}{\href{https://doi.org/10.1063/1.1676739}{J.\
  Chem.\ Phys.}}} \textbf{\bibinfo{volume}{55}}, \bibinfo{pages}{4213--4229}
  (\bibinfo{year}{1971}).

\bibitem{ballenegger-jcp-2005}
\bibinfo{author}{V. Ballenegger} \& \bibinfo{author}{J.-P. Hansen}.
\newblock \emph{\bibinfo{journal}{\href{https://doi.org/10.1063/1.1845431}{J.\
  Chem.\ Phys.}}} \textbf{\bibinfo{volume}{122}}, \bibinfo{pages}{114711}
  (\bibinfo{year}{2005}).

\bibitem{bonthuis-prl-2011}
\bibinfo{author}{D.~J. Bonthuis}, \bibinfo{author}{S. Gekle} \&
  \bibinfo{author}{R.~R. Netz}.
\newblock
  \emph{\bibinfo{journal}{\href{https://doi.org/10.1103/PhysRevLett.107.166102}{Phys.\
  Rev.\ Lett.}}} \textbf{\bibinfo{volume}{107}}, \bibinfo{pages}{166102}
  (\bibinfo{year}{2011}).

\bibitem{bonthuis-l-2012}
\bibinfo{author}{D.~J. Bonthuis}, \bibinfo{author}{S. Gekle} \&
  \bibinfo{author}{R.~R. Netz}.
\newblock
  \emph{\bibinfo{journal}{\href{https://doi.org/10.1021/la2051564}{Langmuir}}}
  \textbf{\bibinfo{volume}{28}}, \bibinfo{pages}{7679--7694}
  (\bibinfo{year}{2012}).

\bibitem{loche-prl-2019}
\bibinfo{author}{P. Loche}, \bibinfo{author}{A. Wolde-Kidan},
  \bibinfo{author}{A. Schlaich}, \bibinfo{author}{D.~J. Bonthuis} \&
  \bibinfo{author}{R.~R. Netz}.
\newblock
  \emph{\bibinfo{journal}{\href{https://doi.org/10.1103/PhysRevLett.123.049601}{Phys.\
  Rev.\ Lett.}}} \textbf{\bibinfo{volume}{123}}, \bibinfo{pages}{049601}
  (\bibinfo{year}{2019}).

\bibitem{schlaich-prl-2016}
\bibinfo{author}{A. Schlaich}, \bibinfo{author}{E.~W. Knapp} \&
  \bibinfo{author}{R.~R. Netz}.
\newblock
  \emph{\bibinfo{journal}{\href{https://doi.org/10.1103/PhysRevLett.117.048001}{Phys.\
  Rev.\ Lett.}}} \textbf{\bibinfo{volume}{117}}, \bibinfo{pages}{048001}
  (\bibinfo{year}{2016}).

\bibitem{jalali-pre-2020}
\bibinfo{author}{H. Jalali} \emph{et~al.}
\newblock
  \emph{\bibinfo{journal}{\href{https://doi.org/10.1103/PhysRevE.102.022803}{Phys.\
  Rev.\ E}}} \textbf{\bibinfo{volume}{102}}, \bibinfo{pages}{022803}
  (\bibinfo{year}{2020}).

\bibitem{stark-cpr-2026}
\bibinfo{author}{P. St\"ark} \emph{et~al.}
\newblock
  \emph{\bibinfo{journal}{\href{https://doi.org/10.1063/5.0315836}{Chem.\
  Phys.\ Rev.}}} \textbf{\bibinfo{volume}{7}}, \bibinfo{pages}{011319}
  (\bibinfo{year}{2026}).
\newblock arXiv:\eprint{2512.07548}.

\bibitem{egberts-ebj-1994}
\bibinfo{author}{E. Egberts}, \bibinfo{author}{S.~J. Marrink} \&
  \bibinfo{author}{H.~J.~C. Berendsen}.
\newblock
  \emph{\bibinfo{journal}{\href{https://doi.org/10.1007/BF00180163}{Eur.\
  Biophys.\ J.}}} \textbf{\bibinfo{volume}{22}}, \bibinfo{pages}{423--436}
  (\bibinfo{year}{1994}).

\bibitem{naumowicz-electrochim-2006}
\bibinfo{author}{M. Naumowicz}, \bibinfo{author}{A.~D. Petelska} \&
  \bibinfo{author}{Z.~A. Figaszewski}.
\newblock
  \emph{\bibinfo{journal}{\href{https://doi.org/10.1016/j.electacta.2006.03.038}{Electrochim.\
  Acta}}} \textbf{\bibinfo{volume}{51}}, \bibinfo{pages}{5024--5028}
  (\bibinfo{year}{2006}).

\bibitem{muzio-langmuir-2020}
\bibinfo{author}{M.~D. Muzio}, \bibinfo{author}{R. Millan-Solsona},
  \bibinfo{author}{J.~H. Borrell}, \bibinfo{author}{L. Fumagalli} \&
  \bibinfo{author}{G. Gomila}.
\newblock
  \emph{\bibinfo{journal}{\href{https://doi.org/10.1021/acs.langmuir.0c02251}{Langmuir}}}
  \textbf{\bibinfo{volume}{36}}, \bibinfo{pages}{12963--12972}
  (\bibinfo{year}{2020}).

\bibitem{petrov-aca-2006}
\bibinfo{author}{A.~G. Petrov}.
\newblock
  \emph{\bibinfo{journal}{\href{https://doi.org/https://doi.org/10.1016/j.aca.2006.01.108}{Anal.\
  Chim.\ Acta}}} \textbf{\bibinfo{volume}{568}}, \bibinfo{pages}{70--83}
  (\bibinfo{year}{2006}).

\bibitem{harland-pre-2010}
\bibinfo{author}{B. Harland}, \bibinfo{author}{W.~E. Brownell},
  \bibinfo{author}{A.~A. Spector} \& \bibinfo{author}{S.~X. Sun}.
\newblock
  \emph{\bibinfo{journal}{\href{https://doi.org/10.1103/PhysRevE.81.031907}{Phys.\
  Rev.\ E}}} \textbf{\bibinfo{volume}{81}}, \bibinfo{pages}{031907}
  (\bibinfo{year}{2010}).

\bibitem{thomas-epl-2021}
\bibinfo{author}{N. Thomas} \& \bibinfo{author}{A. Agrawal}.
\newblock
  \emph{\bibinfo{journal}{\href{https://doi.org/10.1209/0295-5075/134/68003}{Europhys.\
  Lett.}}} \textbf{\bibinfo{volume}{134}}, \bibinfo{pages}{68003}
  (\bibinfo{year}{2021}).

\bibitem{torbati-rmp-2022}
\bibinfo{author}{M. Torbati}, \bibinfo{author}{K. Mozaffari},
  \bibinfo{author}{L. Liu} \& \bibinfo{author}{P. Sharma}.
\newblock
  \emph{\bibinfo{journal}{\href{https://doi.org/10.1103/RevModPhys.94.025003}{Rev.\
  Mod.\ Phys.}}} \textbf{\bibinfo{volume}{94}}, \bibinfo{pages}{025003}
  (\bibinfo{year}{2022}).

\bibitem{dimova-sm-2007}
\bibinfo{author}{R. Dimova} \emph{et~al.}
\newblock \emph{\bibinfo{journal}{\href{https://doi.org/10.1039/B703580B}{Soft
  Matter}}} \textbf{\bibinfo{volume}{3}}, \bibinfo{pages}{817--827}
  (\bibinfo{year}{2007}).

\bibitem{dimova-sm-2009}
\bibinfo{author}{R. Dimova} \emph{et~al.}
\newblock \emph{\bibinfo{journal}{\href{https://doi.org/10.1039/B901963D}{Soft
  Matter}}} \textbf{\bibinfo{volume}{5}}, \bibinfo{pages}{3201--3212}
  (\bibinfo{year}{2009}).

\bibitem{thomas-sm-2022}
\bibinfo{author}{N. Thomas} \& \bibinfo{author}{A. Agrawal}.
\newblock
  \emph{\bibinfo{journal}{\href{https://doi.org/10.1039/D2SM00740A}{Soft
  Matter}}} \textbf{\bibinfo{volume}{18}}, \bibinfo{pages}{6437--6442}
  (\bibinfo{year}{2022}).

\bibitem{kirkwood-jcp-1939}
\bibinfo{author}{J.~G. Kirkwood}.
\newblock \emph{\bibinfo{journal}{\href{https://doi.org/10.1063/1.1750343}{J.\
  Chem.\ Phys.}}} \textbf{\bibinfo{volume}{7}}, \bibinfo{pages}{911--919}
  (\bibinfo{year}{1939}).

\bibitem{neumann-mp-1983}
\bibinfo{author}{M. Neumann}.
\newblock
  \emph{\bibinfo{journal}{\href{https://doi.org/10.1080/00268978300102721}{Mol.\
  Phys.}}} \textbf{\bibinfo{volume}{50}}, \bibinfo{pages}{841--858}
  (\bibinfo{year}{1983}).

\bibitem{klauda-jpcb-2010}
\bibinfo{author}{J.~B. Klauda} \emph{et~al.}
\newblock \emph{\bibinfo{journal}{\href{https://doi.org/10.1021/jp101759q}{J.\
  Phys.\ Chem.\ B}}} \textbf{\bibinfo{volume}{114}},
  \bibinfo{pages}{7830--7843} (\bibinfo{year}{2010}).

\bibitem{abraham-zenodo-2024}
\bibinfo{author}{M. Abraham} \emph{et~al.}
\newblock
  \href{https://doi.org/10.5281/zenodo.11148638}{\bibinfo{title}{{GROMACS
  2024.2 Manual}}} (\bibinfo{year}{2024}).

\bibitem{jorgensen-jcp-1983}
\bibinfo{author}{W.~L. Jorgensen}, \bibinfo{author}{J. Chandrasekhar},
  \bibinfo{author}{J.~D. Madura}, \bibinfo{author}{R.~W. Impey} \&
  \bibinfo{author}{M.~L. Klein}.
\newblock \emph{\bibinfo{journal}{\href{https://doi.org/10.1063/1.445869}{J.\
  Chem.\ Phys.}}} \textbf{\bibinfo{volume}{79}}, \bibinfo{pages}{926--935}
  (\bibinfo{year}{1983}).

\bibitem{berendsen-cpc-1995}
\bibinfo{author}{H. Berendsen}, \bibinfo{author}{D. {van der Spoel}} \&
  \bibinfo{author}{R. {van Drunen}}.
\newblock
  \emph{\bibinfo{journal}{\href{https://doi.org/10.1016/0010-4655(95)00042-E}{Comp.\
  Phys.\ Commun.}}} \textbf{\bibinfo{volume}{91}}, \bibinfo{pages}{43--56}
  (\bibinfo{year}{1995}).

\bibitem{feng-jctc-2023}
\bibinfo{author}{S. Feng}, \bibinfo{author}{S. Park}, \bibinfo{author}{Y.~K.
  Choi} \& \bibinfo{author}{W. Im}.
\newblock
  \emph{\bibinfo{journal}{\href{https://doi.org/10.1021/acs.jctc.2c01246}{J.\
  Chem.\ Theory Comput.}}} \textbf{\bibinfo{volume}{19}},
  \bibinfo{pages}{2161--2185} (\bibinfo{year}{2023}).

\end{thebibliography}

\end{document}